\documentclass
[aps,prc,twocolumn,showpacs,showkeys,amsmath,floatfix,superscriptaddress]
{revtex4-1}
\usepackage{graphicx}
\usepackage{mathptmx}                
\usepackage{dcolumn}                 
\usepackage{bm}                      

\def\be{\begin{equation}}
\def\ee{\end{equation}}
\def\bea{\begin{eqnarray}}
\def\eea{\end{eqnarray}}

\begin{document}

\title{Comparative study of neutron and nuclear matter\\
with simplified Argonne nucleon-nucleon potentials}

\author{M. Baldo}
\affiliation{
INFN Sezione di Catania, Dipartimento di Fisica,
Via Santa Sofia 64, I-95123 Catania, Italy}

\author{A. Polls}
\affiliation{
Departament d'Estructura i Constituents de la Mat\`eria,
Universitat de Barcelona, E-08028 Barcelona, Spain}

\author{A. Rios}
\affiliation{
Department of Physics, Faculty of Engineering and Physical Sciences, 
University of Surrey, Guildford, Surrey GU2 7XH, United Kingdom}

\author{H.-J. Schulze}
\affiliation{
INFN Sezione di Catania, Dipartimento di Fisica,
Via Santa Sofia 64, I-95123 Catania, Italy}

\author{I. Vida\~na}
\affiliation{
Centro de F\'isica Computacional, Department of Physics,
University of Coimbra, PT-3004-516 Coimbra, Portugal}

\begin{abstract}
We present calculations of the energy per particle of pure neutron and symmetric
nuclear matter with simplified Argonne nucleon-nucleon potentials for
different many-body theories. 
We compare critically the Brueckner-Hartree-Fock results to other formalisms,
such as the Brueckner-Bethe-Goldstone expansion up to third order, 
Self-Consistent Green's Functions, 
Auxiliary Field Diffusion Monte Carlo,
and Fermi Hyper Netted Chain. 
We evaluate the importance of spin-orbit and tensor correlations in the equation
of state and find these to be important in a wide range of densities. 
\end{abstract}


\maketitle

\section{Introduction}

The properties of homogeneous nuclear and neutron matter at high density
play a crucial role in the determination of the structure of neutron star
interiors \cite{Shapiro1983}.
Terrestrial nuclei provide little input to constrain the equation of state
(EOS) under the extreme conditions of density and isospin asymmetry within
neutron stars.  A potentially safe way to obtain the EOS thus
relies on microscopic many-body calculations based on realistic
nucleon-nucleon (NN) interactions.  Over the years, several many-body
approaches have been developed to describe neutron star (and nuclear)
interiors.  The different approaches might have very different physical
foundations, but their final result is generally the same: a prediction for
the density dependence of the energy per particle in neutron or symmetric
nuclear matter.  Here we want to progress in our understanding of the EOS
by comparing quantitatively the results provided by different many-body
approaches.  Similar benchmark calculations have taken place within the
few-body community and have provided vital insight into the approximation
schemes at play \cite{kamada2001}.

The full operatorial structure of current high-quality NN potentials is
presently too sophisticated for some state-of-the-art many-body schemes.
Simplified versions of these potentials are therefore useful for
benchmarking purposes.  In particular, we will use a family of simpler
versions of the widely used Argonne $V_{18}$ potential \cite{Wiringa1995}.
The $V'_8$, $V'_6$, and $V'_4$ potentials \cite{Wiringa2002}
are built by removing operatorial components of the interaction, while the
remaining terms are readjusted (as indicated by the prime) in order to
preserve as many lowest-order phase shifts and deuteron properties as
possible.  These potentials have been used in calculations of both infinite
matter and nuclei \cite{Gandolfi2007b}.  A natural question then arises:
how well can these truncated potentials replace the original $V_{18}$?

To clarify this and other issues, we will first analyze in detail the
properties of the family of Argonne potentials by inspecting their phase
shifts in different partial waves.  We will also examine the deuteron
properties as predicted by these interactions.  We will then study the EOS
within the Brueckner-Hartree-Fock (BHF) many-body approach
\cite{baldo1999}, which can handle straightforwardly all the
variants of the Argonne potential.  The BHF results are particularly
insightful, because the total energy can be directly connected to the
partial wave expansion and therefore to the microscopic properties of the
in-medium NN interaction.  We will also compare, when possible, the BHF
results with other many-body approaches \cite{muther2000,dickhoff2008},
namely the Brueckner-Bethe-Goldstone approach up to third order in the
hole-line expansion (BBG)
\cite{song1998}, the Self-Consistent
Green's Function method (SCGF)
\cite{Rios2008,Soma2006,Dickhoff2004}, the Auxiliary
Field Diffusion Monte Carlo (AFDMC) \cite{gandolfi2009}, the
Green's Function Monte Carlo (GFMC) \cite{carlson2003}, and the Fermi Hyper
Netted Chain (FHNC) \cite{lovato2011}.  The comparisons should
be helpful in quantifying theoretical uncertainties with respect to the
EOS.  As we shall see, the symmetric nuclear matter predictions are
particularly susceptible to the missing spin-orbit correlations in
simplified potentials.

This work complements and extends previous investigations of some of the
authors \cite{baldo2004,bombaci2005}.  Our basic conclusions are related to
the spin-orbit and tensor components of the NN interaction and NN in-medium
correlations.  Let us stress, in addition, that we do not attempt to obtain
a ``realistic" description of the neutron or symmetric nuclear matter EOS.
Our goal is solely to compare constructively the results obtained with
different two-body potentials and many-body methods.  A detailed
calculation of the nuclear EOS would need to consider additional effects,
particularly three-body forces \cite{Li2008}, which are beyond the scope of
the present work.

The paper is organized as follows.  In Secs.~II and III we show the phase
shifts and the deuteron properties predicted by the different Argonne
potentials.  Section IV is devoted to the BHF results of nuclear matter,
both symmetric matter (SM) and neutron matter (NM), whereas in Sec.~V we
compare the results obtained with the different many-body approaches.
Finally, our conclusions are presented in Sec.~VI.

\section{Phase shifts}

The strong interaction part of the Argonne $V_{18}$ potential can be
expressed as a sum of 18 operators,
\begin{equation}
  V_{ij} = \sum_{p=1,18} v_p(r_{ij}) O_{ij}^p \:.
\label{eq:eq1}
\end{equation}
The first 14 operators are associated to the spin, isospin, tensor,
spin-orbit, and quadratic spin-orbit components of the nuclear force:
\begin{eqnarray}
 O_{ij}^{p=1,\ldots,14} &=& 
 1,\; \bm\tau_i\cdot\bm\tau_j,\; \bm\sigma_i\cdot\bm\sigma_j,\; 
 (\bm\sigma_i\cdot\bm\sigma_j)(\bm\tau_i\cdot\bm\tau_j), 
\nonumber \\ && 
 S_{ij},\; S_{ij} (\bm\tau_i\cdot\bm\tau_j), 
\nonumber \\ && 
 \bm L \cdot \bm S,\;  
 \bm L \cdot \bm S (\bm\tau_i\cdot\bm\tau_j),
\nonumber \\ && 
 L^2,\; L^2(\bm\tau_i\cdot\bm\tau_j),\; L^2(\bm\sigma_i\cdot\bm\sigma_j),\; 
 L^2 (\bm\sigma_i\cdot\bm\sigma_j)(\bm\tau_i\cdot\bm\tau_j), 
\nonumber \\ && 
 (\bm L \cdot \bm S)^2,\; (\bm L \cdot \bm S)^2 (\bm\tau_i\cdot\bm\tau_j) \:.
\label{e:v}
\end{eqnarray}
The four additional operators,
\begin{equation}
 O_{ij}^{p=15,\ldots,18} = 
 T_{ij},\; T_{ij}(\bm\sigma_i\cdot\bm\sigma_j),\;
 T_{ij}S_{ij},\; ({\tau_z}_i+{\tau_z}_j) \:,
\end{equation}
where 
$T_{ij}= 3 {\tau_z}_i{\tau_z}_j - \bm\tau_i\cdot\bm\tau_j$ 
is the iso-tensor operator, break charge independence.
The radial functions that multiply each operator are adjusted by fitting
experimental data on two-body scattering phase-shifts as well as deuteron
properties.

A family of simplified Argonne NN interactions has been devised to quantify
the evolution of nuclear spectra with increasingly sophisticated NN
interactions \cite{Wiringa2002}.  A given simplified version,
$V'_n$, is constructed by: i) eliminating the operatorial structure with
$p>n$ and ii) refitting the remaining radial functions to reproduce as many
properties of the original interaction as possible.  Thus, $V'_8$ includes
all operators up to the spin-orbit term, but misses the components
proportional to $L^2$, $(\bm L \cdot \bm S)^2$, $T_{ij}$, and
${\tau_z}_i+{\tau_z}_j$.  Similarly, $V'_6$ is a NN force without
spin-orbit (or further, $p>6$) couplings.  Finally, $V'_4$ does not even
have a tensor coupling, but has been refitted to reproduce the binding
energy of the deuteron.

Realistic (or microscopic) NN interactions should fulfill a minimum set of
requirements.  In particular, realistic potentials are built to reproduce
the Nijmegen database \cite{Stoks1993} (which contains a full set of NN
elastic scattering phase shifts up to energies of about 350 MeV) with an
accuracy of $\chi^2/N_\text{data} \sim 1$.  Only potentials that fulfill
this condition should be used as input to the so-called {\it ab initio}
many-body schemes, that aim at providing a {\em first principles}
description of the EOS of neutron and symmetric matter.  While Argonne
$V_{18}$ is, by all means, a realistic interaction, the Argonne family of
simpler versions will necessarily violate this condition and thus they will
become, in some sense, increasingly ``less realistic".  In particular, one
expects the reproduction of the phase shifts of high partial waves to be
deteriorated as the operatorial structure of the interaction is simplified.
One of our aims is to explore to which extent this deterioration has an
impact on the EOS.  This is particularly important in view of the fact that
some approaches, at present, are limited to simplified forms of the Argonne
family of potentials.  In particular, the last generation of FHNC results
has been computed with interactions up to $V'_8$ for both SM and NM.  For
ADFMC, the EOS of NM (SM) is only available with $V'_8$ ($V'_6$).
Diagrammatic approaches, such as BHF, BBG, or SCGF, have less limitations
with respect to the structure of the original NN interactions.  We will
therefore provide results for Argonne $V_{18}$ for these approaches.
Within a given many-body approximation, calculations with $V_{18}$, $V'_8$,
$V'_6$, and $V'_4$ should provide an indication of the importance of the
missing operatorial components for the in-medium properties.

\begin{figure*}[t]
\includegraphics[width=\linewidth,clip]{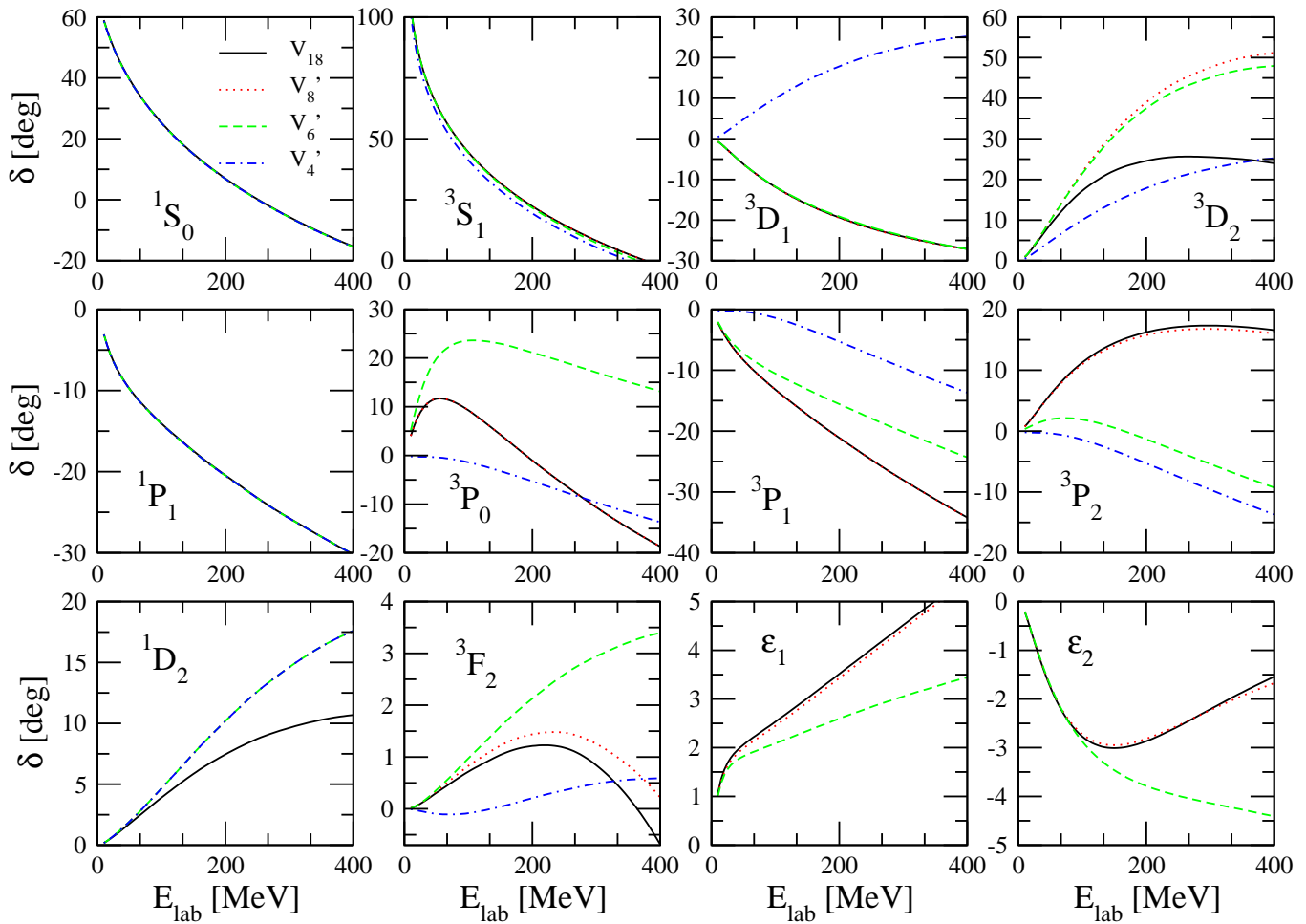}
\caption{(Color online)
NN phase shifts for different potentials as a function of the energy in the
laboratory $E_\mathrm{lab}$.
The solid lines represent the reference Argonne $V_{18}$ results,
whereas dotted, dashed, and dashed-dotted lines correspond to
$V'_8$, $V'_6$, and $V'_4$, respectively.}
\label{f:phase}
\end{figure*}

Fig.~\ref{f:phase} shows the phase shifts of the lowest partial waves given
by the different Argonne potentials \cite{Wiringa2002}.
$V_{18}$, $V'_8$, and $V'_6$ agree, by construction, in the $^1S_0$,
$^3S_1$, $^1P_1$, and $^3D_1$ partial waves.  The over-simplistic $V_4'$
potential, however, does not have a tensor coupling, thus yielding zero
mixing angles $\epsilon_1$ and $\epsilon_2$.  The $^3D_1$ phase-shift is
also badly reproduced for this potential, with an opposite sign relative to
the other potentials.  This is a direct consequence of the readjustment of
the potential to bind the deuteron with the $S$ channel only
\cite{Wiringa2002}.  In the $^3D_1$ wave, the readjustment is such that it
changes the nature of the interaction from repulsive (negative phase-shift)
to attractive (positive phase-shift).

For $L=1$, further discrepancies between phase shifts appear.  All
potentials reproduce the $^1P_1$ phase-shift correctly, as the spin-orbit
or tensor components are not active.  Substantial differences, however,
show up already in the $^3P_{0,1,2}$ waves.  In these channels, the $V'_6$
and $V_4'$ potentials deviate from the experimentally fitted $V_{18}$
results.  In particular, the important $^3P_2$ wave is grossly
misrepresented with the $V_6'$ potential.  In fact, this potential provides
no correction in the $S=1,T=1$ partial waves for the missing spin-orbit
components and can hardly be considered realistic for this reason.

Similarly large discrepancies are observed for the phase shifts of the
$^1D_2$ and $^3D_2$ partial waves.  In the $S=0$ channel, $V'_6$
and $V'_4$ are identical and provide a too large phase-shift compared to
$V_{18}$ and $V'_8$ (which lie on top of each other in the plot). In the
$S=1$ channel, visible differences show up also between $V_8'$ and
$V_{18}$.  For the $L=3$ phase shifts, we have chosen to show the $^3F_2$
channel, where, again, substantial and visible differences appear between
$V'_4$, $V'_6$, $V'_8$, and $V_{18}$.

The lowest right panels of Fig.~\ref{f:phase} show the mixing parameters
$\epsilon_1$ and $\epsilon_2$ of the $^3S_1$-$^3D_1$ and $^3P_2$-$^3F_2$
partial waves, respectively.  On the one hand, the $V'_8$ interaction
reproduces very well the behavior of both parameters as given by the full
$V_{18}$, in spite of the slightly different $^3F_2$ phase shift.  On the
other hand, $V_6'$ is only able to account for the low energy behavior
($E_\textrm{lab} \le 50$ MeV) of $\epsilon_1$ and $\epsilon_2$.  $V_4'$
lacks a tensor component and therefore the mixing parameters associated to
it are zero.

Let us note, for further reference, that the BHF calculations are performed
with partial waves up to $J = 8$.  The SCGF results presented here have
been obtained for up to $J=4$ in the $T$-matrix and up to $J=8$ in the
Hartree-Fock self-energy.  For the BBG 3-hole calculations, all
contributions have $J \leq 5$ and the convergence has been carefully
tested.  The differences in the $J>1$ partial waves that we have just
highlighted will therefore have an impact on the EOS.  As a matter of fact,
even phase-shift-equivalent potentials might predict different EOS due to
their different (and physically unconstrained) off-shell structures.
Naturally, one would expect these differences to be small at low densities,
where the physics is mainly dominated by $L=0$ components.  As the density
increases, however, the differences in the higher partial waves start to
show up.  In non-perturbative diagrammatic calculations, the effects of
high partial waves can be fed back to low momenta due to the
self-consistency procedure.  Likewise, relatively small differences in the
phase shifts can have a substantial impact on the predictions of the EOS at
densities even close to saturation.

\section{Deuteron properties}

\begin{table*}[t]
\caption{
Deuteron $D$-state probability $P_D$, quadrupole moment $Q_d$ (in fm$^2$), 
total binding energy, kinetic and potential energy,
and their decomposition in partial waves,
for different potentials.
All energies are given in MeV.
}
\begin{ruledtabular}
\begin{tabular}{c|dd|ddd|ddddd}
Force & 
 P_D\;(\%) & Q_d
      & E & T & V & T_S &T_D &V_S&V_D&2V_{SD} \\
\hline
$V_{18}$ &5.78&0.27&-2.24&19.86&-22.10&11.30&8.56& -3.95&0.77&-18.91\\
$V'_8$   &5.78&0.27&-2.24&19.86&-22.10&11.30&8.56& -3.95&0.77&-18.91\\
$V'_6$   &5.33&0.27&-2.24&18.70&-20.94&11.38&7.32& -4.68&1.38&-17.64\\
$V'_4$   &0.00&0.00&-2.24&11.41&-13.65&11.41&0.0 &-13.65&0.0 &  0.0 \\
$\tilde V_6$
         &4.64& 0.30 & -1.46&14.96&-16.42& 9.10&5.86& -3.43&1.14&-14.14\\
\end{tabular}
\end{ruledtabular}
\label{t:deut}
\end{table*}

Even though the deuteron only explores the NN potential in the
$^3S_1$-$^3D_1$ partial waves, the analysis of the contributions of the
different waves and operatorial components of the interaction provides a
useful insight into the structure of the interaction \cite{polls1998}.  We
summarize the information related to the deuteron in Table~\ref{t:deut}.
The first column gives the $D$-state probability computed with different NN
interactions.  Although $P_D$ is not an observable, it provides an
indication of the relative importance of the tensor coupling of the
potential.  By construction, the $D$-state probability of $V'_8$ is the
same as that of $V_{18}$, $P_D=5.78\;\%$.  For $V'_6$ the probability
decreases by less than $10\;\%$.  As expected, $P_D=0$ for $V'_4$, whose
deuteron is a pure $S$-wave state.  We also include the results of an
additional potential, $\tilde V_6$, obtained by removing, without any
readjustment, the spin-orbit components from $V_8'$.  For $\tilde V_6$,
$P_D$ is reduced by almost $20\;\%$, indicating the importance of the
spin-orbit components [$p=7,8$ in Eq.~(\ref{e:v})] for the $^3D_1$ wave and
thus the ground state of the deuteron.  Similar statements hold for the
quadrupole moment of the deuteron, which is an observable.  As such, $Q_d$
is reproduced, by construction, with the original $V_{18}$ force and the
refitted $V'_8$ and $V'_6$ forces.  The lack of tensor components in
$V'_4$, however, implies a zero value of $Q_d$.  The non-refitted
interaction $\tilde V_6$ yields a value which is about $10\;\%$ larger than
the experimental one.

In columns 3, 4, and 5, we report the binding energy of the deuteron and
its decomposition into kinetic and potential terms.  All the refitted
potentials reproduce by construction the total binding energy, $E=-2.24$
MeV.  Note that this is the binding energy obtained only with the strong
interaction components of the potential, i.e., when the small
electromagnetic terms are omitted.  These repulsive electromagnetic terms
shift the binding energy to the true experimental value of $E=-2.22$ MeV
\cite{Wiringa2002}.  It is also relevant to note that the charge-dependent
terms of $V_{18}$ ($p=15,\ldots,18$), described in terms of an iso-tensor
operator, have no contribution in the iso-singlet deuteron state.

It is well known that the deuteron binding energy results from a
cancellation between a large positive kinetic and a large negative
potential energy.  For $V_{18}$ these amount to $T=19.86$ MeV and
$V=-22.10$ MeV, respectively.  By construction, $V_8'$ reproduces the same
values as $V_{18}$.  For $V_6'$ there is a small variation in $T$ and $V$
due to the fact that the $^3S_1$ and $^3D_1$ partial waves are not exactly
identical to those of the $V_{18}$ and $V_8'$ potentials.  In particular,
the kinetic energy decreases by about 1 MeV.  $V_4'$ is also able to
reproduce the total binding of the deuteron, but with much smaller kinetic
(and therefore less negative potential) energies.  In contrast, the
potential $\tilde V_6$ looses binding energy and also produces very
noticeable differences for the kinetic and potential energies.  In other
words, a straightforward elimination of the spin-orbit components, without
further readjustments, has large effects for the binding energy.
 
It is also illustrative to separate the contributions of the
$^3S_1$ and $^3D_1$ states to the total kinetic and potential energies.
Assuming that the deuteron is a properly
normalized combination of the $^3S_1$ and $^3D_1$ partial waves,
we define the contributions
of the $S$ and $D$ states to the kinetic energy, 
$T_S = \langle {^3S_1} | T | {^3S_1} \rangle$ and
$T_D = \langle {^3D_1} | T | {^3D_1} \rangle$, 
and to the potential energy,
$V_S = \langle {^3S_1} | V | {^3S_1} \rangle$ and
$V_D = \langle {^3D_1} | V | {^3D_1} \rangle$.
The latter also receives a contribution from the $^3S_1$-$^3D_1$ mixing, 
$V_{SD}=\langle {^3S_1} | V | {^3D_1} \rangle$.
These contributions are listed, for the different potentials,
in columns 6 to 10 of Table~\ref{t:deut}.

For $V_{18}$, $V'_8$, and $V'_6$, the largest contribution to the potential
energy actually comes from the mixing term, $V_{SD}$.  This accounts for
more than $85 \%$ of the final value of the potential energy.  As already
mentioned above, for $V_4'$ the deuteron is a pure $S$-wave state, and
therefore $T_D$, $V_D$, and $V_{SD}$ vanish and the binding energy is
obtained by accumulating a lot of attraction in $V_S$.  In spite of the
fact that the spin-orbit components of $V_{18}$ and $V_8'$ act explicitly
only in the $^3D_1$ partial wave, when these are eliminated without
readjustments in $\tilde V_6$, all the contributions to the binding energy
(and not only $V_D$) are altered.  This is due to the fact that, due to the
tensor coupling, the deuteron is obtained non-perturbatively from a
combination of $S$, $D$, and mixing matrix elements.  One can therefore say
that when the spin-orbit component is not taken into account, a large
change is induced in the wave function of the deuteron.  In a wider
picture, these results illustrate how the elimination of operatorial
components can have a relatively large impact in the binding energy of
nuclear systems. The changes and differences induced by such elimination
become more apparent when different channels are analyzed separately.
While the refitting procedure in the $V'_n$ family of potentials seems to
cure most deficiencies in the case of the deuteron, no such analogous
procedure has been implemented in infinite matter. Consequently, one
expects that even refitted potentials have a significant influence on the
EOS of the infinite system.
   
\begin{table}[t]
\caption{
Contribution of different components
of the potential to the binding energy of the deuteron. 
All energies are given in MeV.
See text for details.
}
\begin{ruledtabular}
\begin{tabular}{c|cccc}
Force & 
Central & Tensor &  Spin-orbit & $L^2$  \\
\hline
$V_{18}$ &-4.45 & -16.62 & -3.75 & 2.72 \\
$V'_8$  &-4.45 & -16.62 & -1.02 & 0.00 \\
$V'_6$  &-5.25 & -15.69 & 0.00 & 0.00 \\ 
$V'_4$  & -13.65 & 0.00 & 0.00 & 0.00 \\
$\tilde V_6$ &-3.84 &-12.58 & 0.00 & 0.00 
\end{tabular}
\end{ruledtabular}
\label{t:deutpot}
\end{table}

Alternatively, one can obtain a quantitative estimation of the different
components of the potential by examining their expectation values in the
ground-state wave function of the deuteron.  We have grouped the 18
components in 4 different sets: the first four operators ($p=1,\ldots,4$),
the tensor components $S_{ij}$ ($p=5,6$), the spin-orbit components $\bm L
\cdot \bm S$ and $( \bm L \cdot \bm S )^2$ ($p=7,8$ and $p=13,14$), and the
quadratic orbital angular momentum components $L^2$ ($p=9,\ldots,12$).  The
group of charge-dependent terms, $p=15,\ldots,18$, does not contribute to
the deuteron, as explained above.  For $V_{18}$, the results of this
decomposition are presented in the first row of Table~\ref{t:deutpot}.  As
expected, the largest contribution corresponds to the tensor component.
All contributions are attractive, except that proportional to $L^2$, which
is slightly repulsive.  The spin-orbit contribution is non-negligible and
amounts to $17\;\%$ of the total potential energy.

The second row of Table~\ref{t:deutpot} shows the results for $V_8'$.  The
total potential energy, i.e., the sum of all the components, is the same as
for $V_{18}$.  However, since the contribution of the first six operators
yield the same result as $V_{18}$ and the quadratic orbital angular
momentum contribution is zero, the contribution of the spin-orbit terms is
reduced to $-1.02$ MeV.  In other words, the absence of the repulsive $L^2$
component is compensated by a smaller attractive contribution of the
spin-orbit components, which decreases to $4.6\;\%$ of the total potential
energy.  Since these spin-orbit effects are more important for $V_{18}$ than
for $V_8'$, any technical difficulties associated to the many-body
treatment of the spin-orbit components in $V_8'$ might artificially neglect
contributions which are important in the case of the real deuteron, coming
from the full interaction.

For $V_6'$ only the first six operators contribute (row 3 in
Table~\ref{t:deutpot}).  These components are about $1$ MeV different from
those in $V_{18}$ and produce a slightly smaller total potential energy
compared to the full results.  As we have discussed previously, for $\tilde
V_6$ there is an important loss of binding energy.  It is interesting to
notice the difference between the spin-orbit contribution provided by
$V_8'$ ($-1.02$ MeV) and the expectation value of the $\bm L \cdot \bm S$
component of $V_8'$ in the ground-state wave function provided by $\tilde
V_6$, which amounts to $-0.58$ MeV.  Obviously, this difference is due to
the different wave functions generated by the two potentials.  The wave
function associated to $\tilde V_6$ does not contain the spin-orbit
correlations that are generated by solving the Schr\"odinger equation with
$V_8'$.  Therefore, the evaluation of the expectation value of the $\bm L
\cdot \bm S$ component of $V_8'$ in the wave function provided by $\tilde
V_6$ gives a poor estimate of the spin-orbit contribution of $V_8'$.  This
suggests that it would be necessary to incorporate the spin-orbit
correlations in the wave function to recover the full contribution.  The
deuteron results illustrate effectively the non-negligible significance of
spin-orbit contributions to the binding energy.

\section{Nuclear EOS in the BHF approach}
\label{s:bhf}

We now proceed to compare the results for the EOS of SM and NM obtained
within the BHF approach for the family of simplified Argonne potentials
described above.  The BHF approach represents the lowest order within the
BBG expansion
\cite{baldo1999,song1998}.
In this formalism, the ground-state energy of infinite matter is computed
from a diagrammatic expansion, which is regrouped according to the number
of independent hole-lines.  Within the BHF approach, the energy is given by
the sum of only two-hole-line diagrams, including the effect of two-body
correlations through the in-medium two-body scattering $G$-matrix.  This
takes into account the effect of the Pauli principle on the scattered
particles and the in-medium potential felt by each nucleon.

As shown in Refs.~\cite{song1998}, the
contribution to the energy from three-hole-line diagrams (which account for
the effect of three-body correlations) is minimized when the so-called
continuous prescription \cite{Jeukenne1976} for the single-particle
potential is adopted.  This is a strong indication of the convergence of
the hole-line expansion and we have adopted this prescription in our BHF
calculations.  We would also like to mention that the $G$-matrix has been
calculated both in $r$- and $k$-space with two independent numerical codes
and an excellent agreement has been found at all densities.  We also remind
the reader that no three-body forces are included in our calculations.

\begin{figure}[t]
\includegraphics[width=80mm,clip]{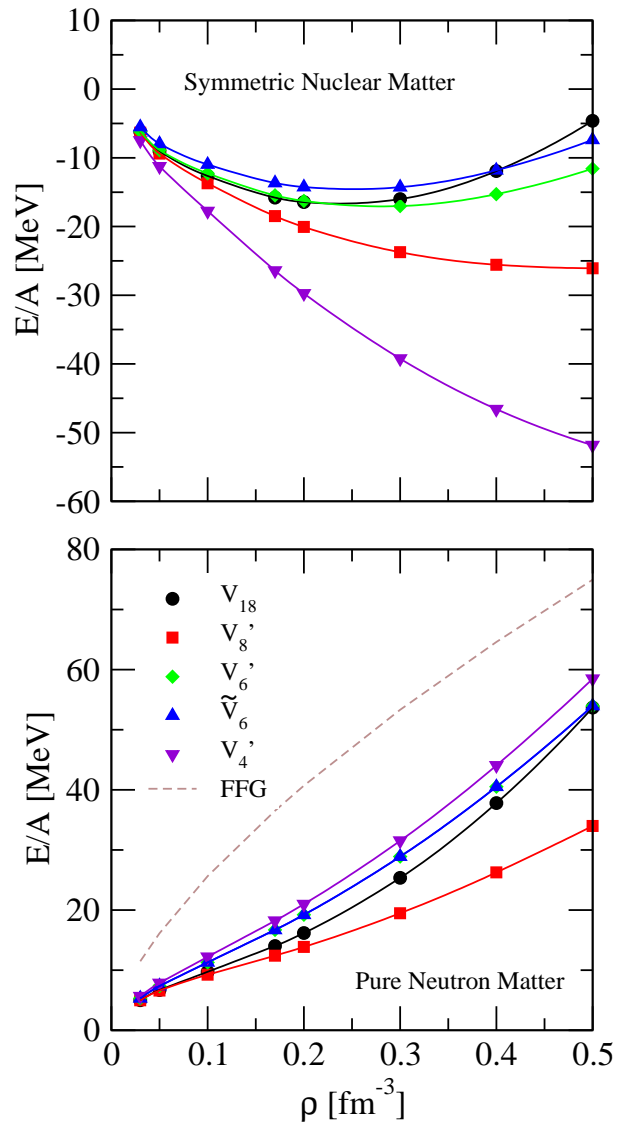} 
\caption{(Color online)
Energy per particle for nuclear (upper panel) and neutron (lower panel)
matter as a function of density calculated in the BHF approach
with the different Argonne potentials.}
\label{f:bhf}
\end{figure}

In general, due to their different off-shell behavior, phase-shift
equivalent potentials might lead to different saturation curves of nuclear
matter.  Within BHF, the saturation points of phase-shift equivalent
interactions lie on the so-called ``Coester band," which provides an
empirical correlation between the saturation energy and density
\cite{li2006}.  Quite obviously, potentials with different phase-shifts
will also predict different saturation properties for infinite matter.  We
shall see in the following that a ``Coester-like" correlation holds,
however, for the different members of the simplified Argonne family of
interactions.

\begin{table}[t]
\setlength{\tabcolsep}{1pt}
\caption{
Energies per particle (in MeV) for SM and NM
at different densities and for different interactions
calculated within the BHF approximation.}
\medskip
\begin{ruledtabular}
\begin{tabular}{d|dd|dd|dd|dd}
     & \multicolumn{2}{c|}{$V_{18}$} & \multicolumn{2}{c|}{$V_8'$}
     & \multicolumn{2}{c|}{$V_6'$}   & \multicolumn{2}{c}{$V_4'$} \\ 
\hline
 \multicolumn{1}{r|}{$\rho\;(\text{fm}^{-3})$}
 & \multicolumn{1}{c}{SM} & \multicolumn{1}{c|}{NM}
 & \multicolumn{1}{c}{SM} & \multicolumn{1}{c|}{NM}
 & \multicolumn{1}{c}{SM} & \multicolumn{1}{c|}{NM}
 & \multicolumn{1}{c}{SM} & \multicolumn{1}{c}{NM} \\
\hline
0.03 &  -6.2&   5.0 &  -6.3 &  4.9 &  -6.1 &   5.3 &  -7.4 &   5.6 \\
0.05 &  -9.0 &   6.7 &  -9.3 &  6.6 &  -8.8 &   7.3 & -11.2 &   7.8 \\
0.10 & -12.6 &   9.7 & -13.7 &  9.2 & -12.2 &  11.2 & -17.7 &  12.2 \\
0.17 & -15.7 &  14.0 & -18.4 & 12.4 & -15.4 &  16.7 & -26.4 &  18.2 \\
0.20 & -16.4 &  16.1 & -20.0 & 13.8 & -16.2 &  19.2 & -29.7 &  20.9 \\
0.30 & -16.0 &  25.3 & -23.7 & 19.4 & -17.0 &  28.9 & -39.2 &  31.5 \\
0.40 & -11.9 &  37.8 & -25.5 & 26.2 & -15.2 &  40.5 & -46.5 &  44.0 \\
0.50 & -4.6  &  53.6 & -26.0 & 33.9 & -11.5 &  53.9 & -51.8 &  58.4 \\
\end{tabular}
\end{ruledtabular}
\label{t:bhf}
\end{table}

In Fig.~\ref{f:bhf}, we compare the total energies of SM (upper panel) and
NM (lower panel) within the BHF approach for the different Argonne NN
forces \cite{baldo1999}.  Several striking features arise from
these comparisons.  To begin with, one immediately notices the relatively
large differences between the $V_{18}$ (circles) and the $V_8'$ (squares)
results in both cases.  This is rather surprising, especially in view of
the close agreements between phase shift and deuteron results presented
earlier.  In both SM and NM, $V_8'$ predicts a much more attractive EOS
compared to $V_{18}$.  In contrast, the $V_6'$ results (diamonds) are
fairly close to the $V_{18}$ calculations, in spite of their relatively
different phase shifts and deuteron predictions.  The $V_4'$ results are
relatively reasonable for NM, where the $^3SD_1$ channel is absent, while
they are clearly unrealistic for SM.  As a matter of fact, symmetric matter
does not even saturate before $\rho=0.5$ fm$^{-3}$ for this potential.  For
a more detailed insight, we refer the reader to Table~\ref{t:bhf}, where
the energies of SM and NM at different densities are listed for the Argonne
family of potentials.

Another relevant question associated to the spin-orbit coupling arises when
comparing the results predicted by the $V_6'$ and $\tilde V_6$ potentials.
$V_6'$ has been constructed specifically by modifying only the central
potential in the $T=0,\, S=1$ states \cite{Wiringa2002}.  As
a consequence, both potentials are identical in the $T=1$ partial waves and
therefore they predict the same EOS for pure NM (see lower panel of
Fig.~\ref{f:bhf}).  The effect of correcting the central $T=0, \, S=1$ term
is only seen in SM, and corresponds to a rather density-independent gain of
binding of a few MeV with respect to the $\tilde V_6$ results (see upper
panel of Fig.~\ref{f:bhf}).

The BHF approach is almost unique within all the many-body approaches, in
the sense that the total energy of the system can be linked to the partial
wave expansion of the in-medium NN interaction.  This provides an
interesting insight into the impact of different partial waves on the EOS
and has motivated us to use BHF as a reference calculation.  An analysis of
the different partial wave contributions to the total energy at
$\rho=0.3\;\text{fm}^{-3}$ is listed in Table~\ref{t:pw} for SM and NM.  In
SM, the largest difference between the $V_{18}$ and $V_8'$ results arises
from the $^3SD_1$ wave ($4.8$ MeV), even though the phase shifts in these
coupled channels are practically identical.  We have verified that this
effect is actually due to the different single-particle potentials.  The
differences in higher-order partial waves thus influence, via the
single-particle states, the lowest partial waves.  Other substantial
contributions to the difference between $V_{18}$ and $V_8'$ stem from the
$^1S_0$ ($1.3$ MeV), the $^1D_2$ ($0.8$ MeV), and the $^3D_2$ ($0.9$ MeV)
states.

Somewhat surprisingly, the total energy of $V_6'$ is actually closer to
that of $V_{18}$, even though the individual partial wave contributions are
rather different.  For instance, the $^3PF_2$ energy
difference is very large ($-12.3$ MeV), in agreement with the unrealistic
phase shifts of $V_6'$ in this channel.  This large discrepancy, however,
is cancelled by opposite (and also relatively large) differences in the
$^3P_0$ and $^3P_1$ channel.  We specify the sum of all $^3P$ states in the
fifth row of Table~\ref{t:bhf}.  This shows that, in the case of $V'_6$,
all $^3P$ waves amount to a difference of only $-3.3$ MeV in the total
energy.  Together with the opposite-sign differences in the $^3SD_1$ ($3.5$
MeV) and $^3D_2$ channels ($0.8$ MeV), the total energy eventually becomes
close to the $V_{18}$ value.

As we have just discussed, the binding energies (and saturation curves) of
SM obtained with the $V_8'$ and the $V_6'$ potentials are rather different.
These large differences are associated to the phase-shift non-equivalence
of the two interactions.  Note, in particular, that the largest differences
between the two sets of results manifest themselves in the channels where
the two potentials are actually not equivalent (i.e., where $V_6'$ has been
modified with respect to $V_8'$).  The different mixing of the $^3SD_1$
channels induces a difference of $1.3$ MeV.  Once again, even though the
$^3P$ individual contributions are rather different, the overall sum leads
to a relatively small difference of $4.1$ MeV.  This is the largest
contribution to the total binding energy difference of $6.9$ MeV.  Since
the energies associated to $V_8'$ are more attractive, a higher saturation
density is found in comparison with $V_{18}$ and $V_6'$.  We conclude that
an intricate series of compensations leads to closer agreement between
$V_6'$ and $V_{18}$ than between $V_8'$ and $V_{18}$. This is clearly
misleading in view of the insufficient phase shift reproduction of $V_6'$
compared to $V_8'$.

\squeezetable
\begin{table}[t]
\caption{
Partial wave decomposition of the binding energy per particle (in MeV)
of SM and NM at $\rho = 0.3\;\text{fm}^{-3}$ for different potentials.
The total sum comprises partial waves up to $J=8$.
The row in brackets contains the partial sum of the
$^3P_0$,$^3P_1$,$^3PF_2$ states.}
\medskip
\setlength{\tabcolsep}{1pt}
\begin{ruledtabular}
\begin{tabular}{@{}d@{\hskip-5mm}|dddd|dddd}
       & \multicolumn{4}{c|}{SM}    & \multicolumn{4}{c}{NM}     \\
\text{State}&V_{18}&V_8'&V_6'& V_4' & V_{18}& V_8' & V_6' & V_4' \\
\hline
T=1  &&&&&&&&\\
^1S_0  & -21.9 &-23.2 &-22.4 &-22.1 & -20.4 &-22.4 &-20.8 &-20.5 \\
^3P_0  &  -5.0 & -5.1 & -9.1 &  0.6 &  -4.2 & -4.4 &-11.4 &  1.4 \\
^3P_1  &  20.2 & 19.9 & 15.6 &  1.8 &  31.5 & 30.8 & 22.6 &  4.2 \\
^3PF_2 & -16.5 &-16.6 & -4.2 &  3.1 & -25.4 &-25.7 & -3.6 &  7.1 \\
(~^3P_*  &  -1.3 & -1.8 &  2.3 &  5.5 &   1.9 & 0.7 &  7.6 & 12.7 ~) \\
^1D_2  &  -5.9 & -6.7 & -6.7 & -6.7 &  -10.0 &-12.1 &-12.1 &-12.0 \\
\hline
T=0 &&&&&&&&\\
^3SD_1 & -20.4 &-25.2 &-23.9 &-43.5 &       &      &      &      \\
^1P_1  &   7.3 &  7.2 &  7.3 &  7.3 &       &      &      &      \\
^3D_2  &  -8.0 & -8.9 & -8.8 & -4.8 &       &      &      &      \\
\hline
\text{All}& -16.0 &-23.7 &-17.0 &-39.2 &  25.3 & 19.4 & 28.9 & 31.5 \\
\end{tabular}
\end{ruledtabular}
\label{t:pw}
\end{table}

A similar issue is observed in the partial wave decomposition of the $V_4'$
results.  The $T=1$ partial waves disagree substantially on a one-by-one
basis (in some cases by more than $10$ MeV), but the overall sum of the
$^3P$ waves differs from the $V_{18}$ results by a smaller number, $-6.8$
MeV.  The most extreme difference is obtained, as expected, in the $^3SD_1$
channel, which is about $20$ MeV more attractive for $V_4'$ than for the
rest of potentials.  Ultimately, it is this extreme additional binding, due
to the lack of partial-wave coupling and the extreme readjustment of the
$^3D_1$ channel, that drives the non-saturating behavior of $V_4'$.

The partial wave differences in NM (see left columns in Table~\ref{t:pw})
also carry interesting information.  On a channel-by-channel basis, the
comparison between $V_{18}$ and $V'_8$ results is more favourable than that
between $V_{18}$ and $V'_6$.  Yet, once again, the discrepancies are
reduced in the overall sum, so that the $V'_6$ final results are closer to
$V_{18}$.  At $\rho=0.3$ fm$^{-3}$, NM is about $10$ MeV more bound with
the $V_8'$ than with $V_6'$.  This difference is entirely due to the
spin-orbit component.  In particular, the phase-shift non-equivalence in
the $^3PF_2$ channel is evidenced by a difference of more than $20$ MeV.
Similarly, the $V'_4$ results for NM are only slightly more repulsive than
the others.  The partial wave decomposition suggests that this is due to a
reshuffling of partial wave contributions which would, separately, deviate
substantially from more realistic results.

Finally, we would like to comment on the specific effect of the spin-orbit
components in a neutron-rich medium.  First, notice that in NM the
differences between $V_6'=\tilde V_6$ and $V_8'$ are only due to the
suppression of the spin-orbit components in $V_8'$, as no other
readjustments are applied.  Clearly the elimination (rather than the
readjustment) procedure for spin-orbit components in the interaction does
not have a small effect in the EOS of NM, as evidenced by the large
differences between the $V'_6$ and $V'_8$ results in Fig.~\ref{f:bhf}.

One might be tempted to attribute the small NM differences between the
$V'_6$ and $V_{18}$ interaction to: i) small spin-orbit components in the
NN interaction or ii) small spin-orbit correlations in the medium.  As we
have seen, however, the partial wave expansion suggests that this is a
rather fortuitous coincidence.  First, both potentials are not truly
phase-shift equivalent.  Second, the in-medium corrections are quite
different, as evidenced by the different partial wave components of the
energy.  All in all, our findings suggest that spin-orbit components should
not be arbitrarily eliminated in \emph{ab initio} calculations.  Extending
the argument to $V'_8$, one could say the same for the remaining missing
operators in the interaction.  Their effect is relevant for the EOS and
needs to be dealt with properly.

A similar reasoning can be extended to SM.  The results provided by $\tilde
V_6$ (without readjusting the potential) and $V_6'$ (with readjustment) are
rather close in that case, indicating that the readjustment in the central
$T=0,\, S=1$ channel is relatively small.  One can therefore say that the
differences between the SM $V_6'$ and $V_8'$ results are mainly due to the
suppression of the spin-orbit component in the interaction itself and they
give rise to rather large corrections.  In other words, the suppression of
the spin-orbit component does not have a small effect on the EOS.  The fact
that the $V_{18}$ and $V'_6$ SM results are relatively close at low
densities is only due to a cancelation effect in different partial waves.
This can hardly be ascribed to a smallness of spin-orbit interactions or
in-medium correlations.  Let us point also out that the situation for the
EOS is different to that of light nuclei, where $V'_6$, $V'_8$, and
$V_{18}$ predict, in the framework of GFMC, somewhat similar spectra and
total energies \cite{Wiringa2002}.  However, it is rather difficult to find
a simple correspondence between the EOS differences presented here and the
GFMC for finite nuclei.

\section{Comparison with other many-body methods}
\label{sec:sec5}

Several approaches have been devised over the decades to treat the nuclear
matter many-body problem.  Critical comparisons between the approaches can
help us learn how correlations, produced by different pieces of the
original NN potentials, are treated within each scheme
\cite{Pandharipande1979,Day1985}.  The ultimate goal of these comparisons
should not be to find a ``good" or a ``bad" EOS, but rather to understand
and quantify the differences among them.  Eventually, a well-founded
comparison might also delineate the theoretical uncertainties of present
generation EOSs.

The only way to carry out a meaningful comparison between many-body
approaches is by starting from the same underlying NN interaction.  While
in principle one would like these to be phase-shift equivalent and as
realistic as possible, some many-body approaches are currently limited by
the operatorial structure of the NN force.  FHNC and AFDMC, in particular,
are coined to treat Argonne-type interactions, involving the sum of
products of local radial functions and operators.  Some parts of the full
$V_{18}$ interaction cannot yet be fully included in these many-body
schemes.  To be able to compare between many-body methods, we will perform
calculations with different approaches with the same family of underlying
NN potentials.  By progressively adding components to the NN force, we also
hope to elucidate the role of such operators in the different many-body
treatments of the EOS.

\begin{figure*}[t]
\includegraphics[width=100mm,clip]{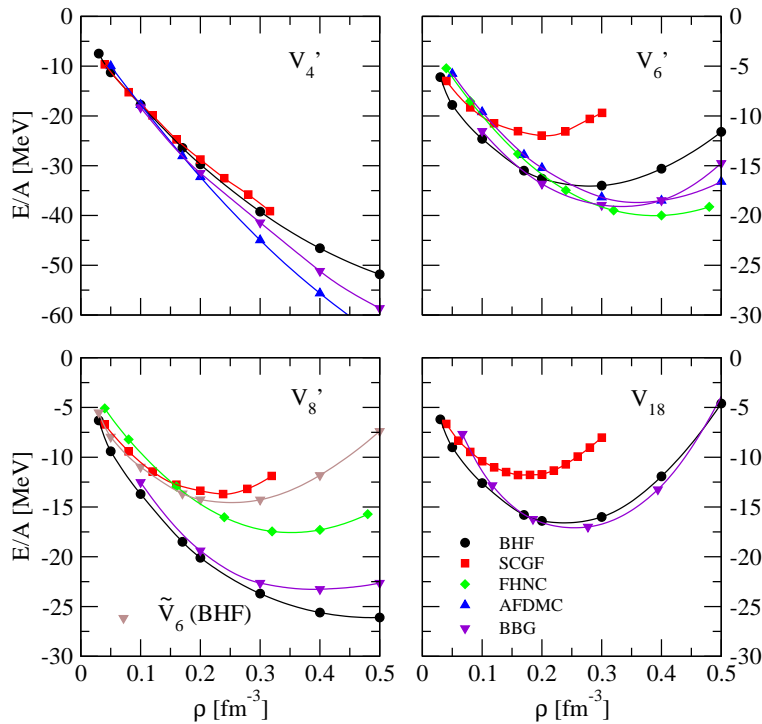}
\caption{(Color online)
Energy per particle of symmetric nuclear matter as a function 
of density calculated with several many-body approaches 
for different Argonne potentials: 
$V_4'$ (upper left), $V'_6$ (upper right),
$V'_8$ (lower left), and $V_{18}$ (lower right).}
\label{f:SM}
\end{figure*}

\begin{figure*}[t]
\includegraphics[width=100mm,clip]{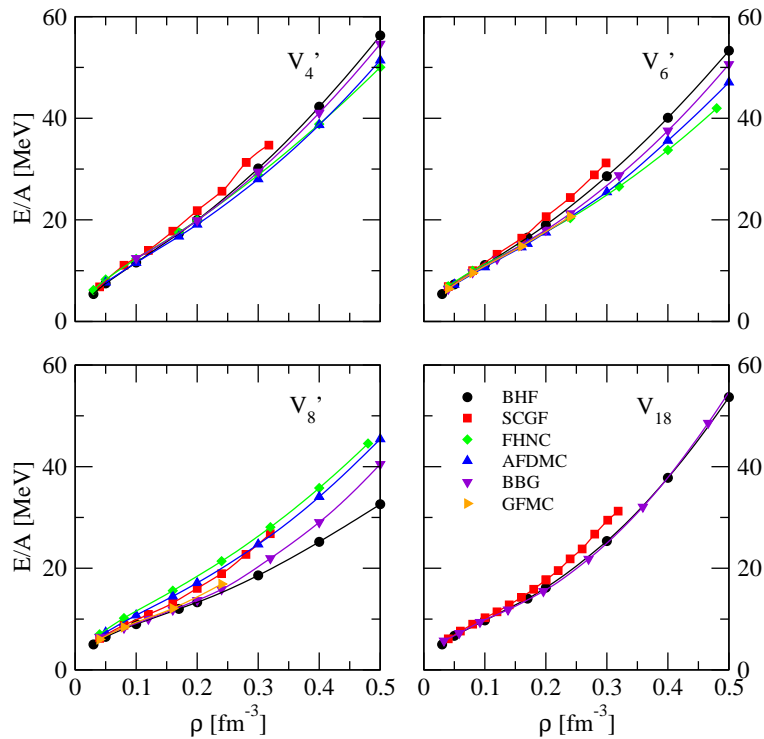}
\caption{(Color online) 
Same as Fig.~\ref{f:SM}, but for neutron matter.}
\label{f:NM}
\end{figure*}

Results are summarized in Figures~\ref{f:SM} and \ref{f:NM}, which
represent the energy per particle as a function of density for both SM and
NM, respectively.  The four panels in each figure give results for
different NN interactions.  Within each panel, BHF results are shown as
circles; SCGF as squares; FHNC as diamonds; AFDMC as up triangles; and BBG
as down triangles.  
In the case of NM we also show  available GFMC (right triangles) results \cite{carlson2003} for the $V'_6$ and $V'_8$ potentials.
Let us first comment the SM results and move later to NM results.

\subsection{Symmetric nuclear matter}

\subsubsection{BBG}

We start by discussing the SM results obtained with the BBG approach at the
three-hole line level. Three-hole line contributions are especially small
when using the continuous prescription for the single-particle spectrum
\cite{song1998}.  Using the gap
prescription, the BBG value converges also to the same results
\cite{Day1985,song1998}.  In this case,
however, the three-hole line contribution is sizeable and therefore the
lowest-order BHF results cannot be taken as a good estimation of the
energy.  We note here that the BBG results shown in this work have been calculated with 
the so-called $K$-matrix \cite{baldo1999} whereas, as it has been said before, the 
BHF ones have been obtained with the $G$-matrix. We have checked, however, that at the BHF level the differences 
between $G$-matrix and $K$-matrix calculations are negligible for both 
SM and NM, and all the potentials. Only a difference of about $2$ MeV is found in 
the case of the $V'_8$ potential for both SM and NM at the highest density considered.

We confirm the good agreement between our continuous choice BHF and the BBG calculations for all the NN interactions considered.
The results differ by
a few MeV at large density, indicating, as already mentioned,
a rather good convergence of the hole-line expansion.
We should mention, however, that  the differences between the BHF and BBG results 
are even smaller when the BHF is done with the $K$-matrix instead of the $G$-matrix.
Note that the
three-hole line contribution can be either attractive or repulsive,
depending on the interaction under consideration.  In particular, it is
minimal for the most realistic potential, $V_{18}$ (lower right panel).
The BBG predictions for $V_8'$ represent minor repulsive corrections to the
BHF results and both lie well below the other approaches.  Note, in
particular, that nuclear matter saturates beyond $0.5$ fm$^{-3}$ in the BHF
case.  As explained earlier, this is due to the partial wave contributions
associated to spin-orbit and tensor forces, which are particularly strong
for this potential.  This is confirmed by comparison with the $\tilde V_6$
results in the same panel, in which these forces are removed and a very
strong reduction of binding is observed.

In spite of the self-consistency process required to calculate the energy,
the BHF approach is not, strictly speaking, thermodynamically
consistent. The Hugenholtz-Van Hove theorem is typically violated by about
20 MeV in BHF calculations \cite{hugenholtz1958,bozek2001}.  The latter
theorem specifies that the chemical potential should equal the
quasi-particle energy at $k_F$.  Within the hole-line expansion, however,
the single-particle spectrum is merely an auxiliary quantity and should
therefore not be taken as a definition of the chemical potential.  A way to
deal with thermodynamical consistency within BHF theory has been discussed
in Ref.~\cite{baldo1999b}.  This should in principle be applicable to BBG
calculations as well.

\subsubsection{SCGF}

Another way to approach the many-body problem is through the SCGF method
\cite{dickhoff2008}.  In this case, a diagrammatic expansion is employed to
solve for the in-medium one-body propagator, rather than for the energy of
the system.  For infinite matter, the method is conventionally applied at
the ladder approximation level.  With respect to the $G$-matrix, the
in-medium ladder interaction presents two major differences.  First,
hole-hole intermediate states are considered in addition to the typical BHF
particle-particle propagators.  Second, these intermediate propagators are
fully dressed, i.e., expressed in terms of spectral functions rather than
through single-particle energies only.

At a formal level, the comparison between the BHF and the SCGF approaches
is not straightforward.  Even though both approaches arise from a
diagrammatic expansion, the infinite subsets of diagrams considered in both
approaches are not the same.  Moreover, the summation procedures are also
somewhat different, with the Dyson equation being used in SCGF to dress all
internal propagators.  Whereas the BHF formalism in the continuous choice
can be derived from the ladder SCGF formalism after a series of
approximations \cite{Frick2004}, this is not the case for the full BBG
expansion.

\begin{figure*}[t]
\includegraphics[width=100mm,clip]{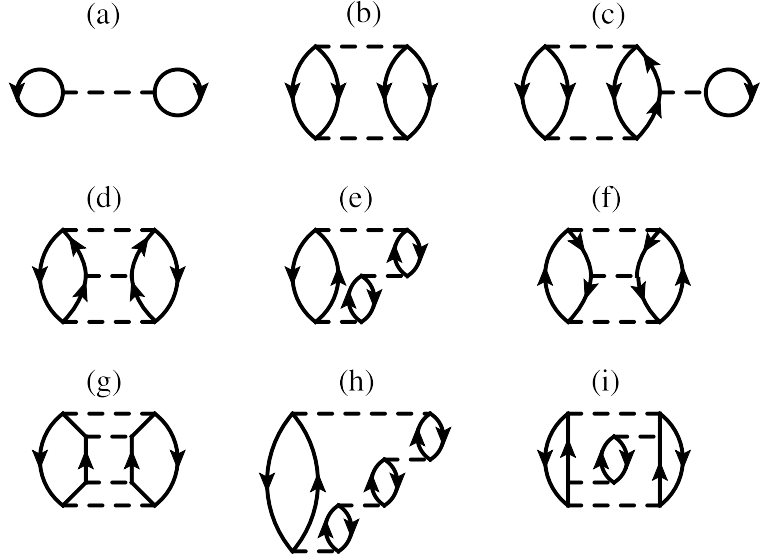}
\caption{
Diagrams in the perturbation expansion of the total energy.
Only direct terms are shown for simplicity.}
\label{f:diags}
\end{figure*}

In principle, if both BBG and SCGF were carried out to all orders, they
should yield identical results.  BBG theory, however, is an expansion in
powers of density (or hole-lines), and the three-hole line results seem to
indicate that it converges quickly.  The error in the SCGF expansion is
more difficult to quantify, as one cannot directly compute (or even
estimate) the values for the diagrams of other structural types.  We show a
few representative diagrams for the perturbation expansion of the energy in
Fig.~\ref{f:diags}.  Dashed (continuous) lines represent interactions
(fermions).  Up to second order, i.e., diagrams (a) and (b), both
approaches include the same diagrammatic contributions.  The only
discrepancies at this level would arise from the (potentially
self-consistent) treatment of internal lines.  Within the SCGF approach,
this would automatically give rise to diagrams like (c), which, in BBG
theory, are considered separately.  At higher orders, ring diagram
iterations, as in diagram (e) or (h), would be included in the three-hole
line BBG expansion, but not in the SCGF ladder resummation.  Similarly,
diagram (i) represents a third-order bubble diagram that is not explicitly
incorporated in the SCGF theory.  Diagram (f) is a typical hole-hole
scattering process included in SCGF and absent in BHF.

It is well established that the differences between SCGF and BHF result in
an overall repulsive effect to the binding energy of both SM
\cite{Dewulf2003,Frick2004,Rios2007} and NM \cite{Rios2008}, which is
mainly due to the inclusion of hole-hole propagation.  Phase space
arguments suggest that this repulsive effect should increase with density.
The dressing of intermediate propagators is capital for the thermodynamical
consistency of the method.  The Hugenholtz-Van Hove theorem is well
fulfilled in these calculations.  At the technical level, we would like to
point out that the SCGF results presented here have been obtained as
zero-temperature extrapolations of finite-temperature calculations.  The
details of the extrapolation procedure will be presented elsewhere
\cite{Rios2012}.  Pure zero-temperature calculations have been obtained
using different numerical techniques by the Ghent \cite{Dewulf2003} and
Krakow \cite{Soma2006} groups.

Compared to BHF, the repulsive effects within the SCGF approach are quite
sizable in SM, especially when the tensor component is considered.  For the
case of $V_{18}$ (lower right panel), for instance, the saturation point
shifts from [$0.25\;\text{fm}^{-3}$,$-16.8\;\text{MeV}$] for BHF to
[$0.17\;\text{fm}^{-3}$,$-11.9\;\text{MeV}$] for SCGF.  While the shift
seems to go towards the right saturation density, the value of the SCGF
saturation energy is quite high.  A similar tendency (lower saturation
density, quite less binding at saturation) is also found for $V_6'$ and
$V_8'$, in agreement with the findings of Ref.~\cite{Dewulf2003}.  The
oversimplified $V_4'$ does not seem to saturate within the SCGF approach
either.  Note that, for this potential, the results are very close to BHF,
which could suggest that, if tensor and spin-orbit correlations are not
present, the many-body problem is under better control (see also the upper
left panel in Fig.~\ref{f:NM}).

In the SCGF approach, the total energy is usually obtained through the
Koltun sum rule \cite{dickhoff2008}.  Even though this gives direct access
to the total true kinetic and potential energies, the sum rule cannot be
explicitly analyzed in terms of partial waves, as we have done for BHF.  It
is therefore difficult to pin down the observed differences to specific
terms in the NN interaction. Alternative ways of computing the energy, as those
suggested in Ref.\ \cite{Soma2006}, could provide such a partial wave decomposition.

\subsubsection{FHNC}

Several approaches based on the variational principle have also been
devised to treat the nuclear matter problem \cite{fabrocini1998}.  These
are rather different from the methods based on non-perturbative sums of
diagrams presented earlier, but the basic principle is rather simple.  The
expectation value of the Hamiltonian,
\begin{eqnarray}
 E = \frac {\langle \Psi_T \mid H \mid \Psi_T \rangle }
 {\langle \Psi_T \mid \Psi_T \rangle} \:,
\end{eqnarray}
in a trial wave function,
\begin{eqnarray}
 \Psi_T = \left(\prod_{i<j}F_{ij}\right)_\text{sym} \phi_\text{Fermi-Gas} \:,
\end{eqnarray}
provides an upper bound to the exact energy.  
The trial wave function is built by means of a correlation operator,
\begin{eqnarray}
 F_{ij} = \sum_p f^p(r_{ij}) O^p_{ij}
\end{eqnarray} 
that describes the correlations induced by the NN interaction.  In
principle, this correlation operator should have the same operatorial
structure as the interaction, see Eq.~(\ref{eq:eq1}).  
In practice, though, this is generally not the case,
since the calculation of the expectation value for the full Hamiltonian is
technically very involved.

The total energy can be evaluated in a diagrammatic cluster expansion with
the aid of the Fermi Hyper Netted Chain/Single Operator Chain (FHNC/SOC)
integral equations \cite{Pandharipande1979}, which sum Meyer-type diagrams
containing up to an infinite number of nucleons.  The sum is, however,
incomplete, as some topologies and operatorial structures are difficult to
include within infinite summations.  For instance, the elementary diagrams
are generally not included in FNHC calculations.  Similarly, the spin-orbit
correlations cannot be chained, and are usually evaluated at the three-body
cluster level.  On top of this variational upper bound to the energy, one
can add second-order perturbative corrections ($\Delta E^\text{CBF}$)
calculated in the framework of the correlated basis function (CBF) theory
\cite{fabrocini1998}.

The FHNC results for SM with $V'_8$ and $V'_6$ are presented in the
corresponding panels of Fig.~\ref{f:SM}.  For our comparisons, we use the
FHNC calculations reported in
Refs.~\cite{lovato2011,lovato_pri}.  There is no clear
tendency with respect to our BHF reference calculations.  Whereas the
$V'_6$ FHNC results are more attractive than the BHF calculations, a
significantly repulsive behavior is observed for $V'_8$.  Consequently, the
saturation points are quite different to those of BHF.  For $V_6'$, the FHNC
results saturate around [$0.39\;\text{fm}^{-3}$,$-20.1\;\text{MeV}$], while
the $V_8'$ calculation saturates at
[$0.35\;\text{fm}^{-3}$,$-17.6\;\text{MeV}$].

If the variational procedure had been exactly performed, the FHNC results
should provide an upper bound to the energy per particle.  However, as
pointed out in Refs.~\cite{lovato2011}, the uncertainties and
difficulties in the treatment of the spin-orbit correlations in the FHNC
formalism have hindered the inclusion of spin-orbit correlations in the
wave function, i.e., the underlying NN interaction and correlation function
are not treated on an equal footing.  A widely debated question is if the
variational character of FHNC results is preserved after the approximations
involved in the calculational procedure.  Recent comparisons between FHNC
and AFDMC results have enlightened this basic question \cite{lovato2011}.
Notice also that the inclusion of 2 particle-2 hole corrections calculated
at second order in the framework of the CBF formalism would give an
additional attraction to the FHNC results \cite{bombaci2005}.

Contrary to what we have observed within BHF, the FHNC results seem to be
quite independent of the underlying interaction.  This has been attributed
to the smallness of the spin-orbit components of the NN interaction as well
as the cancellation of spin-orbit correlations in the in-medium wave
function.  These arguments stem from comparisons with AFDMC results
\cite{lovato2011}.  The impossibility to switch off spin-orbit correlations
in BHF hinders a direct comparison with the $V_8'$ FHNC
calculations, where the spin-orbit correlations have been omitted.  As in
the case of the deuteron, to have a quantitative idea of the contribution
of the spin-orbit components, we report BHF results for the $\tilde V_6$
interaction.  The EOS of SM with $\tilde V_6$ is shown in the lower left
panel of Fig.~\ref{f:SM}.  As explained earlier, the difference between the
$V_8'$ and $\tilde V_6$ results is only caused by the spin-orbit components
in the original interaction.  The differences turn out to be rather large
and therefore this comparsion, although not conclusive, does not support
the idea that switching off spin-orbit correlations produces small changes
in the energy.  In any case, FHNC calculations with a more elaborate
treatment of the spin-orbit correlations would be highly desirable.

\subsubsection{QMC}

Quantum Monte Carlo (QMC) methods are very successful in describing the
ground state of infinite system of bosons, like atomic liquid $^4$He
\cite{Boronat1994}, or fermions, like liquid $^3$He \cite{Casulleras2000}.
Additional efforts have allowed the extension of the QMC method to nuclear
systems, which have more complicated interactions and correlation
structures.  However, the accuracy of QMC in its different versions, be it
AFDMC \cite{gandolfi2009} or GFMC \cite{carlson2003}, is
limited by the fermion sign problem \cite{kalos1984}.  Up to now, the
safest way to deal with this problem in nuclear systems is to keep the
sample walk within a fixed nodal surface.  This is an approximation and,
consequently, a potential limitation of the QMC approach.  AFDMC and GFMC
differ in the way they treat the spin and isospin degrees of freedom.
AFDMC samples the spin-isospin states by introducing Hubbard-Stratonovich
auxiliary fields, whereas GFMC sums them completely.  This fact limits in a
drastic way the number of nucleons that GFMC can consider, generally up to
about 16.  The auxiliary field strategy allows AFDMC to efficiently sample
spin-isospin correlations in systems with a sufficient number of nucleons
($N=114$) to ensure that the finite size corrections are small.  A recent
comparison has demonstrated that both methods give very close results for
neutron drops with $N \leq 16$ \cite{Gandolfi2011}.

In spite of its recent progress, the AFDMC method is still not able to work
with the full $V_{18}$ potential.  Technical problems, again mainly
associated with the spin-orbit structure of the interaction and the trial
wave function, are the reason for the lack of AFDMC results for SM with
potentials containing spin-orbit components ($V_{18}$ or $V_{8}'$).  We
will use the most recent version of the AFDMC results in our comparisons
\cite{gandolfi_pri}.  $V_4'$ results are rather similar to our BHF
reference calculation up to $0.2\;\text{fm}^{-3}$, but the AFDMC
predictions become more attractive as density increases.  The AFDMC results
for SM with $V'_6$, as shown in the upper right panel of Fig.~\ref{f:SM},
are quite close to the FHNC predictions.  Note for instance that AFDMC with
$V_6'$ saturates around [$0.37\;\text{fm}^{-3}$,$-18.8\;\text{MeV}$].
However, the agreement between both approaches should not be taken as a
final consistency check.  Note that these results still present a series of
limitations, particularly those associated to the sign problem.

Let us finally comment on the results obtained with $V_4'$.  As an
unrealistic interaction without a tensor component, $V_4'$ yields an
unrealistic EOS for SM which does not saturate for any of the many-body
approaches discussed here.  This potential, however, can be used for
benchmarking purposes.  As a matter of fact, in spite of its limitations,
the $V_4'$ results are useful, as we find a large degree of agreement among
the different many-body approaches.  This confirms the long accepted
notion that, in the nuclear matter case, the correlations induced by
spin-orbit and tensor components of the interaction are probably 
on the basis of the large differences when comparing different many-body 
methods.  As a matter of fact, the antisymmetrization procedure in NM 
eliminates part of these components (particularly, the tensor coupling) 
and, as we shall see in the following, this results into a much closer 
agreement among all the many-body calculations.

\subsection{Neutron matter}

The NM results for different NN interactions and many-body approaches are
presented in Fig.~\ref{f:NM}.  Overall, it is fair to say that the
differences between the different potentials are significantly smaller for NM than
for SM.  As just mentioned, we attribute this similarity to the fact that,
for a fully isospin-polarized system, the number of active partial waves is
reduced.  This is particularly true for the $^3SD_1$ coupled wave, which is
inactive in this system.  Consequently, the differences between potentials
are further reduced.  Moreover, the lack of tensor correlations presumably
facilitates the solution of the many-body problem.

For $V_4'$ (upper left panel), all reported approaches give quite similar
results.  This supports the idea that, when the potential is just
central, a good agreement between the different many-body techniques is
found.  Let us in particular stress the fact that, up to
$\rho=0.3\;\text{fm}^{-3}$, the results of all the different approaches
fall within a rather narrow window of less than $5$ MeV.

When the tensor component of the force is taken into account, as in $V_6'$
(see upper right panel), the differences up to $\rho = 0.3 $fm$^{-3}$ are
still relatively small.  At this density, we find $e^\text{BHF} = 28.6$
MeV, $e^\text{BBG}=26.9$ MeV, $e^\text{SCGF}= 31.2$ MeV, $e^\text{AFDMC}=
25.5$ MeV, and $e^\text{FHNC}= 25.0$ MeV.  In other words, the inclusion of
the tensor component increases the uncertainty (measured as the spread
predicted by these calculations) in the many-body calculations by $1-2$
MeV.  Note that, however, a broader discrepancy is observed at high
densities, with FHNC lying below all the other approaches.  In general,
SCGF provides the more repulsive results.  As mentioned earlier, $V_6'$ and
$\tilde V_6$ produce the same results for neutron matter because the
readjustment of the potential does not affect the NM partial waves.

Since the results for $V_6'$ and $\tilde V_6$ are the same for NM, the
differences between $V_8'$ and $V_6'$ give a direct measure of the
spin-orbit contributions in the NM channels where it is active.  The
incorporation of the spin-orbit components in $V_8'$ produces an overall
attraction with respect to the $V_6'$ results, apart from FHNC.  The most
attractive variation corresponds to BHF, while the SCGF, FHNC, and AFDMC
approaches stay rather close to their $V_6'$ counterparts at all densities.
Note that also GFMC results seem to have a stronger dependence on the underlying
interaction, and that they are in good agreement with the BBG ones.
At $\rho=0.3\;\text{fm}^{-3}$, we find $e^\text{BHF}=18.3$ MeV,
$e^\text{BBG}=20.4$ MeV, $e^\text{SCGF}=24.8$ MeV, $e^\text{AFDMC}=24.7$
MeV, and $e^\text{FHNC}=26.4$ MeV.  The spread in these results has
increased substantially compared to the simpler potentials.  Again, the
differences increase with density.  Notice, however, that in this
case the FHNC results stay above all the other methods for all densities,
whereas for $V_6'$ this approach provided the most attractive results.

We would like to stress once again that the spin-orbit contribution in the
NN interaction produces a sizeable contribution for the NM BHF EOS with
respect to $V_6'$.  This is in contrast to the small effect observed in the
AFDMC or the FHNC methods.  For these two methods one should notice that
spin-orbit correlations (and also the tensor ones in the AFDMC case) are not 
included in the trial wave function, although the AFDMC allows for their fully generation 
in the diffusion process.

The results for $V_{18}$ are only available for the three perturbative
diagrammatic approaches, BHF, SCGF, and BBG. The agreement between them is
in general rather good.  At $0.3\;\text{fm}^{-3}$, for instance, we find
$e^\text{BHF}=25.4$ MeV, $e^\text{BBG}=25.2$ MeV, and $e^\text{SCGF}=29.4$
MeV.  This represents a relatively small spread in values compared with the
differences we have observed in SM. One should keep in mind that 
these differences increase with density. Therefore, as a general remark we can conclude that in NM the
change of the EOS with respect to the BHF results for both SCGF and BBG is
less important than in SM.  This is particularly true for the $V_4'$ and
$V_6'$ potentials.  For the $V_8'$ potential, larger differences are observed
between these three approaches even in NM.  In all cases, the SCGF method
produces the more repulsive EOS.

\section{Conclusions}

By performing calculations of the energy per particle for nuclear and
neutron matter within different many-body approaches, we have investigated
the properties of the equation of state as obtained with a family of
frequently used Argonne potentials.  The many-body approaches we have used are: the
Brueckner-Hartree-Fock, the Brueckner-Bethe-Goldstone approach up to third
order in the hole-line expansion, the Self-Consistent Green's Function
method, the Auxiliary Field Diffusion Monte Carlo, and the Fermi Hyper
Netted Chain.  We have analyzed critically the origin of 
underlying uncertainties in these calculations.  The subtraction and refitting
procedure of different
operatorial structures, inherent in the family of Argonne NN interactions used in
the present paper, has
provided us with substantial insight.

At the phase shift and deuteron levels, we find that already the $V_6'$
potential produces fairly large deviations relative to the original
$V_{18}$ model.  While at the two-body level $V_8'$ is almost a clone of
$V_{18}$ for $S$ and $P$ partial waves, we have found that, when included
in many-body calculations, it produces relatively large differences.  These
discrepancies are driven by the phase-shift differences in higher partial
waves as well as by off-shell effects.  The $V_8'$ NN interaction should
therefore be regarded critically in high-precision applications.

The overall infinite-matter binding energies obtained with the $V_6'$
interaction are actually closer to the $V_{18}$ results, in spite of the
unsatisfactory reproduction of the $P$-wave phase shifts.  We believe that
this agreement is, however, due to an artificial reshuffling of different
partial wave contributions.  To support this claim, we have presented the
partial wave decomposition of the total energy in BHF calculations, which
has indeed confirmed an overall cancellation of larger differences in the
total energy.  In the $T=1$ channel, this potential is not readjusted for
the missing spin-orbit component, which otherwise produces fairly important
contributions to the binding energy.  It might well be that microscopic
properties, other than the EOS, are also affected by these large
discrepancies.
 
Finally, while the $V_4'$ model is clearly unrealistic and should only be
used for benchmarking purposes, we have found that it is at this level that
the different many-body approaches agree more.  This confirms the long
accepted hypothesis that the tensor (and spin-orbit) components of the
interaction and their in-medium treatment are at the heart of most of the
observed discrepancies.  Overall, one needs to consider these with more
attention before drawing more definite conclusions. Studies like the one
presented here help us in quantifying the uncertainties in state-of-the-art
predictions of the EOS, originating either from the microscopic interaction or
the many-body theory. Only a thorough understanding of these uncertainties 
will allow us to provide a well-founded connection between the physics of neutron
stars, the fundamental strong interaction acting on its constituents, and the many-body
correlations at play.

\begin{acknowledgments}

We thank O. Benhar, S. Gandolfi, A. Lovato, and A. Illarionov
for useful discussions. 
In particular, we thank S. Gandolfi and A. Lovato for providing us with the
AFDMC and the FHNC results, respectively.
This work is partly supported by COMPSTAR, an 
ESF (European Science Foundation) Research Networking Programme;
by FEDER; by the initiative QREN financed by the UE/FEDER through the 
Programme COMPETE under the projects, PTDC/FIS/113292/2009, 
CERN/FP/109316/2009, CERN/FP/116366/2010 and CERN/FP/123608/2011;
by the Consolider Ingenio 2010 Programme CPAN CSD2007-00042,
Grant No. FIS2008-01661 from MEC and FEDER (Spain)
and Grant No. 2009GR-1289 from Generalitat de Catalunya (Spain);
by STFC through an Advanced Fellowship (ST/I005528/1) and Grant ST/J000051;
and by the MIUR PRIN grant 2008KRBZTR.

\end{acknowledgments}


\end{document}